\documentclass[aps,pra,reprint,amsmath,amssymb,noshowpacs,floatfix]{revtex4-1}

\usepackage{graphicx,units}

\newcommand{\ket}[1]{\ensuremath{\vert{#1\rangle}}} 
\newcommand{\bra}[1]{\ensuremath{{\langle #1}\vert}}
\newcommand{\braket}[2]{\ensuremath{{\langle #1}\vert{#2 \rangle}}}
\newcommand{\ketbra}[2]{\ensuremath{|{#1 \rangle}{\langle #2}|}}
\newcommand{\op}[1]{\hat{#1}}

\newcommand{\I}{\text{i}}
\newcommand{\E}{\text{e}}
\providecommand{\abs}[1]{\left\lvert#1\right\rvert}

\usepackage[colorlinks,citecolor=blue,urlcolor=blue,linkcolor=blue,hyperindex,breaklinks]{hyperref}

\begin{document}

\title{Observation of the quantum paradox of separation \\of a single photon from one of its properties}
\author{James M.\ Ashby}
\author{Peter D.\ Schwarz}
\author{Maximilian Schlosshauer}
\thanks{Author to whom correspondence should be addressed. Electronic mail: \texttt{schlossh@up.edu}}
\affiliation{Department of Physics, University of Portland, 5000 North Willamette Boulevard, Portland, Oregon 97203, USA}

%\date{\today}

\begin{abstract} 
We report an experimental realization of the quantum paradox of the separation of a single photon from one of its properties (the so-called ``quantum Cheshire cat''). We use a modified Sagnac interferometer with displaced paths to produce appropriately pre- and postselected states of heralded single photons. Weak measurements of photon presence and circular polarization are performed in each arm of the interferometer by introducing weak absorbers and small polarization rotations and analyzing changes in the postselected signal. The absorber is found to have an appreciable effect only in one arm of the interferometer, while the polarization rotation significantly affects the signal only when performed in the other arm. We carry out both sequential and simultaneous weak measurements and find good agreement between measured and predicted weak values. In the language of Aharonov \emph{et al.} and in the sense of the ensemble averages described by weak values, the experiment establishes the separation of a particle from one its properties during the passage through the interferometer. \\[-.1cm]

\noindent Journal reference: \emph{Phys.\ Rev.\ A\ }\textbf{94}, 012102 (2016), DOI: \href{http://dx.doi.org/10.1103/PhysRevA.94.012102}{10.1103/PhysRevA.94.012102}
\end{abstract}


\maketitle

\section{Introduction}

The concept of quantum weak values obtained from weak measurements was first introduced by Aharonov, Albert, and Vaidman \cite{Aharonov:1988:mz,Duck:1989:uu} and has since become an important experimental tool. It is used, for example, for the amplification of weak detector signals \cite{Ritchie:1991:uu}, high-precision estimation of evolution parameters \cite{Hosten:2008:op,Dixon:2009:ii}, direct measurement of quantum states \cite{Lundeen:2011:ii,Lundeen:2012:rr,Bamber:2014:ee} and geometric phases \cite{Sjoqvist:2006:ii,Kobayashi:2010:az}, the study of quantum paradoxes \cite{Aharonov:2002:aq,Lundeen:2009:pp,Yokota:2009:aa,Resch:2004:yy}, investigation of nonclassical behavior \cite{Mir:2007:uu,Dressel:2011:au,Groen:2013:az}, and measurement of physical quantities with minimum state disturbance \cite{Palacios-Laloy:2010:zp,Kocsis:2011:za}. The weak value of an observable $\op{A}$ represents the average of a series of weak measurements conditioned on a pre- and postselected ensemble \cite{Dressel:2014:uu}. Specifically, for a system in the initial (preselected) state $\ket{\psi}$ and postselected in the state $\ket{\phi}$, the (first-order \cite{Lorenzo:2012:zz,Dressel:2014:uu}) weak value of $\op{A}$ is defined by \cite{Aharonov:1988:mz}
\begin{equation}\label{eq:haavsg7117788aaiuxxh}
\langle \op{A} \rangle_w = \frac{ \bra{\phi}\op{A}\ket{\psi}}{\braket{\phi}{\psi}}.
\end{equation}
Weak values have the precise operational and experimentally relevant meaning of quantifying the change in the detection probability $\abs{\braket{\phi}{\psi}}^2$ caused by a weak intermediate unitary interaction generated by $\op{A}$ \cite{Dressel:2014:uu}.

Recently, weak values have been used to demonstrate a new quantum paradox called a quantum Cheshire cat \cite{Aharonov:2005:ii,Aharonov:2013:im,Matzkin:2013:ee,Bancal:2014:yy,Denkmayr:2014:rr,Atherton:2015:oo,Correa:2015:ii}, named for the apparent separation of the location of a quantum system from one of its physical properties, akin to the separation of the Cheshire cat's grin from the cat in the story of \emph{Alice in Wonderland}. The original proposal by Aharonov \emph{et al.}  \cite{Aharonov:2013:im} used weak measurements on an appropriately pre- and postselected ensemble of photons passing through an interferometer to establish the following situation. If one performs a weak measurement to test for photon presence in one arm of the interferometer, the corresponding weak value always indicates presence in only one arm, while the weak value corresponding to a weak measurement of polarization indicates that the photon's polarization is carried by the other arm. Due to their weakness, both measurements can be carried out simultaneously, leading Aharonov \emph{et al.} \cite{Aharonov:2013:im} to suggest that ``photon polarization may exist where there is no photon at all'' and that ``physical properties can be disembodied from the objects.'' The quantum Cheshire cat provides a fruitful setting for the theoretical and experimental study of weak values and weak measurements. 

The first experimental realization of a quantum Cheshire cat \cite{Denkmayr:2014:rr} used single neutrons and performed the pair of weak measurements sequentially. A subsequent experiment \cite{Atherton:2015:oo} implemented a photonic equivalent with simultaneous weak measurements. However, since it did not use single photons, it merely generated a classical Cheshire cat, in the sense that the experimental results can be explained using classical waves \cite{Atherton:2015:oo}. 

Here we report an experimental realization of a single-photon quantum Cheshire cat based on heralded photons traversing a Sagnac-like interferometer. We realize the weak measurement of presence by inserting weak absorbers (Brewster-angle glass slides) into one arm of the interferometer and measuring the change in photon counts at the exit port of the interferometer. For the weak measurement of polarization, we perform small polarization rotations in one arm and observe the resulting change in the interference pattern \cite{Denkmayr:2014:rr,Atherton:2015:oo}. Our experiment also allows for simultaneous measurements of presence and polarization. 

This paper is organized as follows. Section~\ref{sec:theory-impl} explains the theory of the quantum Cheshire cat in the context of the implementation used in our experiment. Section~\ref{sec:experimental-setup} describes the experimental setup. Section~\ref{sec:results} reports the results of our experiment. We offer concluding remarks in Sec.~\ref{sec:conclusion}.

\section{\label{sec:theory-impl}Theory}

\subsection{\label{sec:weak-measurements}Weak measurements}

Following the approach of Refs.~\cite{Denkmayr:2014:rr,Atherton:2015:oo}, we implement the weak measurement of presence by placing high-transmission absorbers into the path of the photon, and we realize the weak measurement of polarization through a small rotation of the photon's polarization. The weak values then appear as a small change in the detection probabilities \cite{Dressel:2014:uu,Denkmayr:2014:rr}, as follows. Let $\op{\Pi}_1 =\ketbra{1}{1}$ and $\op{\Pi}_2 = \ketbra{2}{2}$ be the projection operators onto the eigenstates $\ket{1}$ and $\ket{2}$ corresponding to the two possible paths 1 and 2 through an interferometer. Suppose we place a weak absorber with coefficient of transmission $T_k$ into path $k$ ($k=1,2$), such that a photon is absorbed with probability $R_k=1-T_k$. (In our experiment, absorption is realized by scattering an incoming photon from the beam.) Then an arbitrary input state $\ket{\psi}$ is changed to 
\begin{equation}\label{eq:hvsguhthgu22}
\ket{\psi} \, \longrightarrow \, \ket{\psi'} = N \left[1- \left(1-\sqrt{T_k}\right) \op{\Pi}_k\right]\ket{\psi},
\end{equation}
where $N$ is a normalization factor that, for weak absorption, we can approximate by $N=1$. The probability of detecting the final (postselected) state $\ket{\phi}$ is
\begin{align}\label{eq:hvsguhthgu222}
\abs{\braket{\phi}{\psi'}}^2 &\approx \abs{\braket{\phi}{\psi}}^2\left[1-2 \left(1-\sqrt{T_k}\right)\,\mathrm{Re}\,\langle \op{\Pi}_k\rangle_w\right]\notag \\ &\approx  \abs{\braket{\phi}{\psi}}^2\left[1-R_k\,\mathrm{Re}\,\langle \op{\Pi}_k\rangle_w\right],
\end{align}
where $\langle \op{\Pi}_k \rangle_w$ is the weak value of $\op{\Pi}_k$ [see Eq.~\eqref{eq:haavsg7117788aaiuxxh}] and the last approximation holds for weak absorption ($R_k=1-T_k \ll 1$). Thus, $\mathrm{Re}\,\langle \op{\Pi}_k\rangle_w$ quantifies the change in the detection probability caused by the absorber. 

Similarly, consider a weak measurement of circular polarization in arm $k$ of the interferometer, represented by the observable $\op{\sigma}_\text{circ} \op{\Pi}_k = (\ketbra{+}{+}-\ketbra{-}{-}) (\ketbra{k}{k}) $ with eigenvalues $0,\pm 1$. Here $\ket{\pm}=\left(\ket{H}\pm \I\ket{V}\right)/\sqrt{2}$ are the eigenstates of the circular-polarization observable $\op{\sigma}_\text{circ}=\left(\begin{smallmatrix}0&-\I\\\I&0\end{smallmatrix}\right)$, where $\ket{H}$ and $\ket{V}$ denote horizontal and vertical linear polarization, respectively. The observable $\op{\sigma}_\text{circ}$ generates the unitary transformation
\begin{equation}
\op{U}(\theta)=\E^{-\I \theta \op{\sigma}_\text{circ}} = \cos\theta -\I \op{\sigma}_\text{circ}\sin\theta = \begin{pmatrix}\cos\theta &-\sin\theta\\ \sin\theta&\cos\theta\end{pmatrix},
\end{equation}
which rotates linear polarization by an angle $\theta$ and describes the action of a half-wave plate oriented at angle $\theta/2$ from the vertical. Up to second order in $\theta$, $\op{U}(\theta)$ implements the state transformation
\begin{align}\label{eq:hvsguhthgu22aa}
\ket{\psi} \, \longrightarrow \, \ket{\psi'} &= \E^{-\I \theta \op{\sigma}_\text{circ} \op{\Pi}_k}\ket{\psi} \notag \\ &\approx \left(1-\I\theta \op{\sigma}_\text{circ} \op{\Pi}_k - \frac{\theta^2}{2} \op{\Pi}_k \right) \ket{\psi},
\end{align}
where we have used that $\left(\op{\sigma}_\text{circ}\op{\Pi}_k\right)^2=\op{\Pi}_k$. Then the probability of detecting the final (postselected) state $\ket{\phi}$ is, up to second order in $\theta$,
\begin{align}\label{eq:hyhyhy}
\abs{\braket{\phi}{\psi'}}^2 &\approx \abs{\braket{\phi}{\psi}}^2\biggl(1+2\theta \,\mathrm{Im}\,\langle \op{\sigma}_\text{circ}\op{\Pi}_k\rangle_w\notag \\ &\qquad -\theta^2 \,\mathrm{Re}\,\langle \op{\Pi}_k\rangle_w + \theta^2 \abs{\langle \op{\sigma}_\text{circ}\op{\Pi}_k\rangle_w}^2\biggr).
\end{align}

\subsection{\label{sec:pre-postselection}Pre- and postselection}

The pre- and postselection procedure for our quantum Cheshire cat is as follows. Consider a horizontally polarized photon passing through a 50--50 beam splitter. The transmitted and reflected beams travel along spatially separated trajectories through the interferometer. The state after the beam splitter is $\ket{\psi_0}=\frac{1}{\sqrt{2}}\left(\ket{1}+\ket{2}\right)\ket{H}$, where $\ket{1}$ and $\ket{2}$ again represent localization in arms~1 and 2 of the interferometer, respectively (we omit phase shifts caused by the beam splitter, since below we introduce a variable relative phase). Both beams travel the same optical distance before being recombined at the beam splitter. A half-wave plate placed in arm~1 changes the polarization from horizontal to vertical, encoding path information in the polarization degree of freedom. We also introduce an adjustable phase shift $\phi$ in arm~2. Then the state $\ket{\psi_0}$ becomes
\begin{equation}\label{eq:hvsguhtddhgu}
\ket{\psi}=\frac{1}{\sqrt{2}}\left(\ket{1}\ket{V}+\E^{\I \phi}\ket{2}\ket{H}\right),
\end{equation}
where $\ket{V}$ denotes vertical polarization. Equation~\eqref{eq:hvsguhtddhgu} is the preselected state in our experiment. 

Applying the weak-measurement scheme described in Sec.~\ref{sec:weak-measurements}, we now place into each arm of the interferometer weak absorbers with coefficients of transmission $T_1$ and $T_2$, and half-wave plates oriented to induce polarization rotations by angles $\theta_1$ and $\theta_2$. Then the state $\ket{\psi}$ in Eq.~\eqref{eq:hvsguhtddhgu} becomes
\begin{align}\label{eq:fv}
\ket{\psi'} &\propto \left[ \sqrt{T_1} \ket{1}\biggl(-\sin\theta_1\ket{H}+\cos\theta_1\ket{V}\right) &\notag \\ &\qquad + \sqrt{T_2} \,\E^{\I \phi} \ket{2}\left(\cos\theta_2\ket{H}+\sin\theta_2\ket{V}\right)\biggr].
\end{align}
The two beams recombine at a beam splitter. By detecting only the horizontally polarized component emerging from output port~1 of the beam splitter, we postselect the state 
\begin{align}\label{eq:post}
\ket{\phi} = \frac{1}{\sqrt{2}}\left(\ket{1}+\ket{2}\right)\ket{H}.
\end{align}

\subsection{Weak values and detection probabilities}

For the pre- and postselected states~\eqref{eq:hvsguhtddhgu} and \eqref{eq:post}, the weak values of $\op{\Pi}_k$ and $\op{\sigma}_\text{circ}\op{\Pi}_k$ are
\begin{align}\label{eq:wk}
\langle \op{\Pi}_1 \rangle_w &= 0, \quad \langle \op{\Pi}_2 \rangle_w = 1, \notag\\ \langle \op{\sigma}_\text{circ}\op{\Pi}_1  \rangle_w &=-\I\E^{-\I \phi}, \quad \langle \op{\sigma}_\text{circ}\op{\Pi}_2 \rangle_w = 0.
\end{align}
For these values, we can use Eqs.~\eqref{eq:hvsguhthgu222} and \eqref{eq:hyhyhy} to calculate the theoretically predicted relative change $\Delta p=\frac{\abs{\braket{\phi}{\psi}}^2-\abs{\braket{\phi}{\psi'}}^2}{\abs{\braket{\phi}{\psi}}^2}$ in the final detection probabilities caused by the intermediate weak measurements, where, as before, $\ket{\psi}$ denotes the preselected state, $\ket{\phi}$ the postselected state, and $\ket{\psi'}$ the preselected state transformed by the weak measurements. For weak measurements of presence in arms 1 and 2, we find
\begin{subequations}
\begin{align}
\Delta p(\text{presence, arm 1}) &\approx 0, \label{eq:p1} \\
\Delta  p(\text{presence, arm 2}) &\approx R_2, \label{eq:p2}
\end{align}
\end{subequations}
Thus, the detection probability remains unchanged if the absorber is placed in arm~1 while it is reduced by a fraction $R_2$ if the absorber is placed in arm~2. This is physically reasonable, since the postselection excludes the state component corresponding to arm~1, and therefore the detection probability (given successful postselection) cannot depend on an absorber present in that arm but will depend linearly on the transmission if the absorber is placed in arm~2. 

For weak measurements of polarization in arms 1 and 2, to leading order in $\theta$ we have
\begin{subequations}
\begin{align}
\Delta p(\text{polarization, arm 1}) &\approx 2\theta_1\cos\phi, \label{eq:p3}\\
\Delta  p(\text{polarization, arm 2}) &\approx 0. \label{eq:p4}
\end{align}
\end{subequations}
Equation~\eqref{eq:p3} describes an interference pattern as a function of the variable phase $\phi$, where the fringe visibility (contrast) is determined by $\theta_1$. Equation~\eqref{eq:p4} shows that a polarization rotation in arm~2 does not affect the detection probability. This is so because even though the polarization rotation will cause this arm to acquire a vertically polarized component that can subsequently interfere with the vertically polarized arm~1, we postselect the horizontal component and therefore no interference fringes will appear. In summary, Eqs.~\eqref{eq:p1}--\eqref{eq:p4} show that, on average, a weak measurement of presence in arm~1 and a weak measurement of polarization in arm~2 will have no effect on the probability of detecting the state $\ket{\phi}$ at the output of the beam splitter. This realizes the quantum Cheshire cat in the sense defined in Refs.~\cite{Aharonov:2013:im,Denkmayr:2014:rr}. 

We may also establish the equivalent result without the use of weak values \cite{Atherton:2015:oo,Correa:2015:ii}, by directly considering the probability $P_H$ that a horizontally polarized photon entering the interferometer will be found to have horizontal polarization at output port~1. In the absence of absorbers ($T_1=T_2=1$) and polarization measurements ($\theta_1=\theta_2=0$), we have $P_{H}= \abs{\braket{\phi}{\psi}}^2 = \frac{1}{4}$, where $\ket{\psi}$ and $\ket{\phi}$ are the pre- and postselected states \eqref{eq:hvsguhtddhgu} and \eqref{eq:post}. In the presence of absorbers and polarization measurements, from Eq.~\eqref{eq:fv} we have
\begin{align}\label{eq:wkpreditt}
  P_H &= \abs{\braket{\phi}{\psi'}}^2 = \abs{\braket{\phi}{\psi}}^2 \biggl[ T_1\sin^2\theta_1 + T_2\cos^2\theta_2 \notag\\ & \quad -2\sqrt{T_1T_2} \cos\theta_2\sin\theta_1 \cos\phi\biggr],
\end{align}
with corresponding fringe visibility
\begin{align}\label{eq:visvis}
V= \frac{V_\text{max}-V_\text{min}}{V_\text{max}+V_\text{min}} = 
\frac{ 2\sqrt{T_1T_2} \cos\theta_2\sin\theta_1 }{ T_1\sin^2\theta_1 + T_2\cos^2\theta_2}.
\end{align}
In the absence of polarization measurements ($\theta_1=\theta_2=0$), 
\begin{align}\label{eq:fhsgvh55}
P_H =T_2\abs{\braket{\phi}{\psi}}^2,
\end{align}
i.e., the detection probability is proportional to the transmission of the absorber in arm~2. This result is in agreement with Eqs.~\eqref{eq:hvsguhthgu222} and \eqref{eq:p2}. Note that no interference pattern is observed because the two arms have orthogonal polarizations. 

Now consider the case of  polarization measurements (i.e., nonzero $\theta_1$ and $\theta_2$), with no absorbers present. For small $\theta_1$ and $\theta_2$, the fringe visibility given by Eq.~\eqref{eq:visvis} becomes 
\begin{align}\label{eq:kjdvghdv}
V \approx  2\sin\theta_1 \approx 2 \theta_1,
\end{align}
and Eq.~\eqref{eq:wkpreditt} gives $P_H \propto 1 - 2 \theta_1 \cos\phi$, in agreement with Eq.~\eqref{eq:p3} derived from the weak value. Thus, the visibility of the interference pattern is predominantly determined by $\theta_1$, i.e., by the weak polarization measurement in arm~1. Finally, in the presence of both weak absorption and weak polarization measurements, the fringe visibility is
\begin{align}\label{eq:visvisrotabs}
V= 2 \sqrt{\frac{T_1}{T_2}} \sin\theta_1 \approx \left(2+T_1-T_2\right)\sin\theta_1,
\end{align}
and Eq.~\eqref{eq:wkpreditt} gives
\begin{align}\label{eq:wkprediff2}
P_H \propto T_2 - 2\sqrt{T_1T_2} \,\theta_1\cos\phi \approx  T_2 - \left(T_1+T_2\right) \theta_1\cos\phi.
\end{align}
This establishes a result analogous to that represented by the weak values in Eq.~\eqref{eq:wk}: For weak measurements, the detection probability is influenced mainly by $T_2$ and $\theta_1$, i.e., by a polarization measurement in arm~1 and a presence measurement in arm~2. In particular, the overall intensity is chiefly determined by $T_2$ but not by $T_1$ (i.e., it is only sensitive to a presence measurement in arm~2), while the visibility of the fringe pattern depends on $\theta_1$ but not on $\theta_2$ (i.e., it is only sensitive to a polarization measurement in arm~1).

\section{\label{sec:experimental-setup}Experimental apparatus}

\begin{figure}
\includegraphics[scale=1.1]{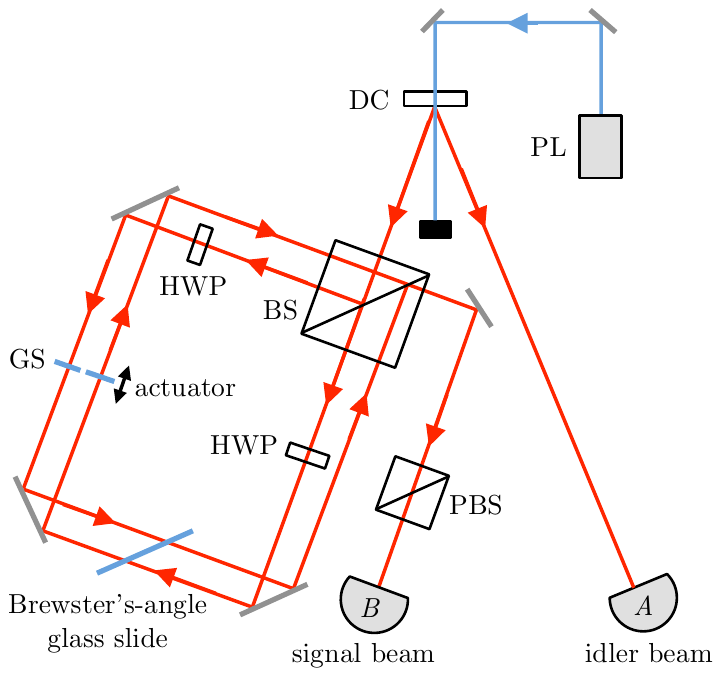}
\caption{\label{fig:setup}(Color online) Experimental apparatus. Pairs of correlated, horizontally polarized 810-nm photons are generated by spontaneous parametric down-conversion in a beta-barium borate crystal (DC) pumped by a 405-nm diode laser (PL). Detection of one photon (the idler) heralds the production of the other photon (the signal). The signal photon travels through a modified Sagnac interferometer with displaced paths implemented with a nonpolarizing beam splitter (BS). Microscope glass slides (GS) are used to adjust the relative path length in the interferometer, where the tilt of one slide is controlled by a motorized actuator. Half-wave plates (HWP) in each path realize the weak polarization measurement through a small rotation of polarization. A microscope glass slide oriented at Brewster's angle implements the weak measurement of presence through its probabilistic rejection of one polarization component. After the two beams are recombined, a polarizing beam splitter (PBS) postselects the horizontally polarized component. Idler and signal photons are captured by fiber-coupling lenses (labeled $A$ and $B$, respectively) and transmitted via multimode fiber-optic cables to single-photon counting modules (not shown).}
\end{figure}

Our experimental apparatus is shown in Fig.~\ref{fig:setup}. A 405-nm, 150-mW laser diode pumps a pair of stacked, 0.5-mm-thick beta-barium borate crystals to produce pairs of 810-nm, polarization-correlated photons through type-I spontaneous parametric down-conversion. The axes of the two crystals are oriented at right angles with respect to each other, which enables generation of polarization-entangled photons. In the present experiment, we do not need such an entangled state and therefore we pump only one of the crystals, producing a photon pair in which both the signal and idler photons are horizontally polarized. We use the idler photon to herald the production of the signal photon and register photons in coincidence between the signal and idler beams to ensure single photons. 

While the original proposal for a quantum Cheshire cat \cite{Aharonov:2013:im} is based on a Mach--Zehnder interferometer, our experiment uses a modified Sagnac interferometer. The signal photon enters a 1-inch nonpolarizing beam splitter away from the center, such that the two paths through the interferometer are displaced with respect to each other while maintaining equal optical path lengths. This separation allows for each path to be manipulated individually. Half-wave plates (HWPs) are introduced into each arm to implement the weak measurement of polarization as discussed in Sec.~\ref{sec:theory-impl}. One of these HWPs has the additional function of changing the polarization from horizontal to vertical, as required for generating the preselected state in Eq.~\eqref{eq:hvsguhtddhgu}.  (In what follows, any angle $\theta_2$ referring to a polarization rotation in arm~2 will not include this initial rotation $\ket{H} \rightarrow \ket{V}$.) To adjust the relative phase $\phi$ between the arms of the interferometer, we insert a 1-mm-thick microscope glass slide into each arm. With both slides initially oriented at right angles to the beam, one of the slides is then tilted about a horizontal axis to change the path length traveled by the photon inside the glass. We vary the tilt angle with a motorized actuator, which allows us to map out an interference pattern as a function of the position of the actuator. 

To weakly measure photon presence, we insert a 1-mm-thick microscope glass slide tilted to Brewster's angle such that both arms of the interferometer pass through it. When the tilt is about a vertical axis, a vertically polarized photon will be reflected (and therefore effectively removed from the beam) with some probability while a horizontally polarized photon remains unaffected. Similarly, a tilt about a horizontal axis partially reflects horizontal polarization while leaving vertical polarization unaffected. If, as in Sec.~\ref{sec:theory-impl}, we label the vertically polarized arm as ``arm~1'' and the horizontally polarized arm as ``arm~2,'' then tilting the slide to Brewster's angle about either the vertical axis or the horizontal axis realizes weak absorption in arms~1 and 2, respectively. In what follows, we refer to these two scenarios as ``filtering arm~1'' and ``filtering arm~2.'' For a glass--air interface, Brewster's angle is $56.3^\circ$ and the calculated reflection probability (reflectivity) is $R=0.148$. Thus, our approach corresponds to placing a weak absorber with transmission $T=1-R=0.852$ into either one of the arms.

The two paths through the interferometer are recombined at the nonpolarizing beam splitter. Only one output is used and sent through a polarizing beam splitter to postselect the transmitted (horizontally polarized) component. The signal and idler photons are captured by fiber-coupling lenses and fed into single-photon counting modules (SPCMs) via multimode fiber-optic cables. Stray photons are removed by 780-nm long-pass filters placed in front of the inputs of the SPCMs. The SPCMs use silicon avalanche photodiodes with a photon detection efficiency of about 30\% at the relevant wavelength of \unit[810]{nm}. Coincidence counting is performed by a field-programmable gate array, with the coincidence window set to around 8\,ns. At each position of the actuator adjusting the relative phase $\phi$, photon counts are taken over a period of \unit[5]{s} before the actuator is moved to the next step.

\section{\label{sec:results}Results}

\subsection{\label{sec:weak-meas-pres}Weak measurement of presence}

\begin{figure}
\includegraphics[scale=.9]{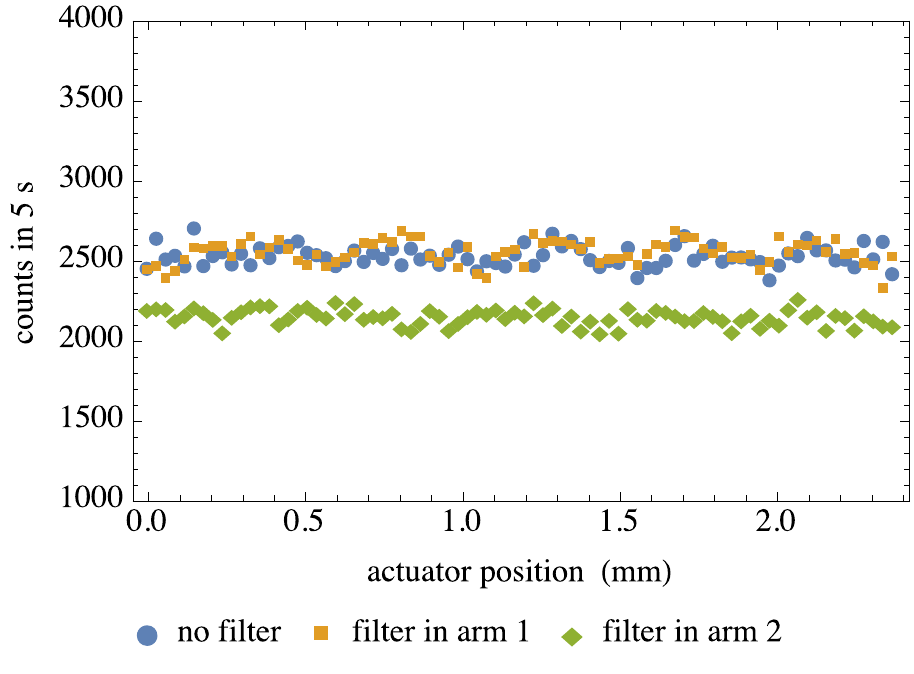}
\caption{\label{fig:filteronly}(Color online) Photon counts (per \unit[5]{s}) when weak measurements of presence are performed inside the interferometer by filtering one of the arms, shown as a function of the position of the actuator that adjusts the relative phase between the two arms of the interferometer. The measurement is implemented by a Brewster-angle glass slide oriented to probabilistically reject either horizontal or vertical polarization. In agreement with theoretical predictions, the count rates are substantially affected only by a measurement of presence in arm~2.}
\end{figure}

We first perform weak measurements of presence without polarization measurements (rotations). Figure~\ref{fig:filteronly} shows the photon counts as a function of actuator position when a Brewster-angle glass slide (the ``filter'') is introduced into the interferometer but no polarization rotations are made ($\theta_1=\theta_2=0$). It is immediately seen that the count rate is substantially affected only if the measurement of presence is carried out in arm~2, establishing the first part of our quantum Cheshire cat. Specifically, when arm~1 is filtered, the photon count averaged over the range of actuator positions is $\langle N_1 \rangle = 2537(8)$, which remains unchanged within the error from the count rate in the absence of the filter, $\langle N_0 \rangle =2526(7)$ (stated uncertainties in $\langle N \rangle$ are standard deviations of the mean, SDM). On the other hand, when arm~2 is filtered, the count average drops to $\langle N_2 \rangle = 2146(6)$, which corresponds to a relative decrease in intensity of $\frac{\langle N_0 \rangle-\langle N_2 \rangle}{\langle N_0 \rangle} = 0.151(8)$ (we use twice the counting-rate SDM for estimating the error to include the effect of small experimental imprecisions in state preparation and filter orientation). This value agrees with the prediction of Eqs.~\eqref{eq:p2} and \eqref{eq:fhsgvh55}, which show that in the absence of polarization measurements, the detection probability is reduced by the Brewster's-angle probability $R=0.148$ for weak absorption in arm~2. Sinusoidal fits of the data give fringe visibilities of less than 2\% without filter and when filtering arm~2, and 3\% for filtering arm~1, indicating that the polarizations in each arm are approximately orthogonal and that the pre- and postselected states in our experiment are indeed close to the ideal pre- and postselected states given by Eqs.~\eqref{eq:hvsguhtddhgu} and \eqref{eq:post}. 

Using Eq.~\eqref{eq:hvsguhthgu222} and the photon-counting averages $\langle N \rangle$, we obtain the experimental weak values $\langle \op{\Pi}_k\rangle_w$ from
\begin{equation}\label{eq:aahvsguhthgu222}
\langle \op{\Pi}_k\rangle_w =\frac{1}{R_k}\frac{\langle N_0 \rangle-\langle N_k \rangle}{\langle N_0 \rangle}, \qquad k=1,2.
\end{equation}
Here we have neglected the imaginary part of $\langle \op{\Pi}_k\rangle_w$, which we justify below. Since $\frac{\langle N_0 \rangle-\langle N_k \rangle}{\langle N_0 \rangle}$ is the measured intensity change in the presence of the absorber and $R_k$ is the calculated reflection probability for a glass slide oriented to filter  arm $k$, one sees that $\langle \op{\Pi}_k\rangle_w$ represents the ratio of measured to predicted absorption. We find the experimental weak values $\langle \op{\Pi}_1\rangle_w = -0.03(4)$ and $\langle \op{\Pi}_2\rangle_w = 1.02(4)$, which agree within the error with the theoretical values $\langle \op{\Pi}_1\rangle_w = 0$ and $\langle \op{\Pi}_2\rangle_w = 1$  [see Eq.~\eqref{eq:wk}] for the ideal pre- and postselected states given by Eqs.~\eqref{eq:hvsguhtddhgu} and \eqref{eq:post}.  Stated uncertainties are estimated from experimental imprecisions in the polarization preparation and postselection, as follows.  We consider the more general pre- and postselected states 
\begin{equation}\label{eq:ttt1}
\ket{\psi}=\frac{1}{\sqrt{2}}\left[\ket{1}\left( \cos\delta_1\ket{V}+\sin\delta_1 \ket{H} \right)+\E^{\I \phi}\ket{2}\ket{H}\right]
\end{equation}
and
\begin{equation}\label{eq:ttt2}
\ket{\phi} = \frac{1}{\sqrt{2}}\left(\ket{1}+\ket{2}\right)\left(\cos\delta_2\ket{H}+ \sin\delta_2\ket{V}\right).
\end{equation}
Hence, we allow for the polarizations in the two arms to be nonorthogonal ($\delta_1 \not =0$) and for the postselection to be not exactly horizontal ($\delta_2\not=0$). The cumulative angle offset $\delta \equiv \delta_1+\delta_2$ measures the deviation from orthogonality in the polarization between arm~1 and the postselection. From Eq.~\eqref{eq:haavsg7117788aaiuxxh}, the corresponding predicted weak values are 
\begin{align}\label{eq:devpostdcc}
\langle \op{\Pi}_1\rangle_w &= \left[1+ \frac{\E^{\I \phi}\cos\delta_2 }{\sin(\delta_1+\delta_2)}\right]^{-1} \approx \delta\E^{-\I \phi}, \notag\\ \langle \op{\Pi}_2\rangle_w &= 1- \langle \op{\Pi}_1\rangle_w,
\end{align}
where the approximation holds for small $\delta_1$ and $\delta_2$. Equation~\eqref{eq:devpostdcc} shows that the imaginary part of the weak value may be neglected provided $\delta$ is small. Due to the small beam separation in the interferometer, our experiment uses edgeless mounts rather than rotation stages for the wave plates, which imposes limitations on the accuracy of reading and setting the wave-plate angles. We estimate this uncertainty in the wave-plate settings to be $1^\circ$,  with a corresponding $2^\circ$ uncertainty in polarization rotation. This $2^\circ$ uncertainty is used in Eq.~\eqref{eq:devpostdcc} (taking $\phi=0$) to estimate the errors in the weak values quoted above. It also translates into a 4\% uncertainty in the measured intensity changes, which is similar to the 3\% uncertainty estimated from the SDM. The observed fringe visibilities of $\lesssim 3\%$ indicate imperfect orthogonality with a corresponding angle offset $\delta_1 \approx \sin^{-1} \left[\frac{1}{2}(0.03) \right]\approx 1^\circ$ [Eq.~\eqref{eq:visvisrotabs}], which is within the range given by the uncertainty in polarization rotation.

\subsection{\label{sec:weak-meas-polar}Weak measurement of polarization}

\begin{figure}

\begin{flushleft}
\hspace{0cm}{\small (a)}
\end{flushleft}

\vspace{-.7cm}

\includegraphics[scale=.9]{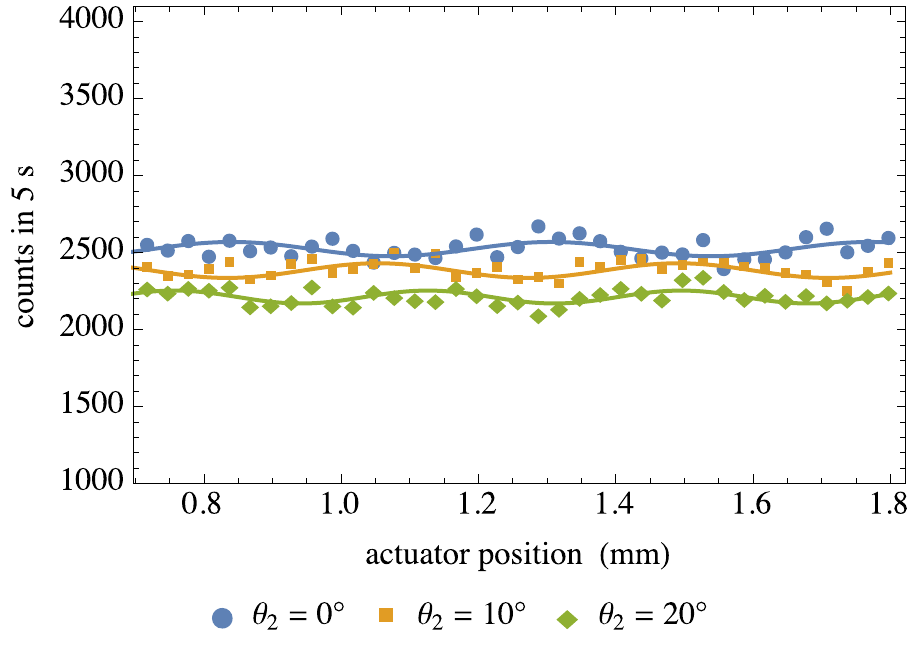}

\vspace{-.2cm}

\begin{flushleft}
\hspace{0cm}{\small (b)}
\end{flushleft}

\vspace{-.7cm}

\includegraphics[scale=.9]{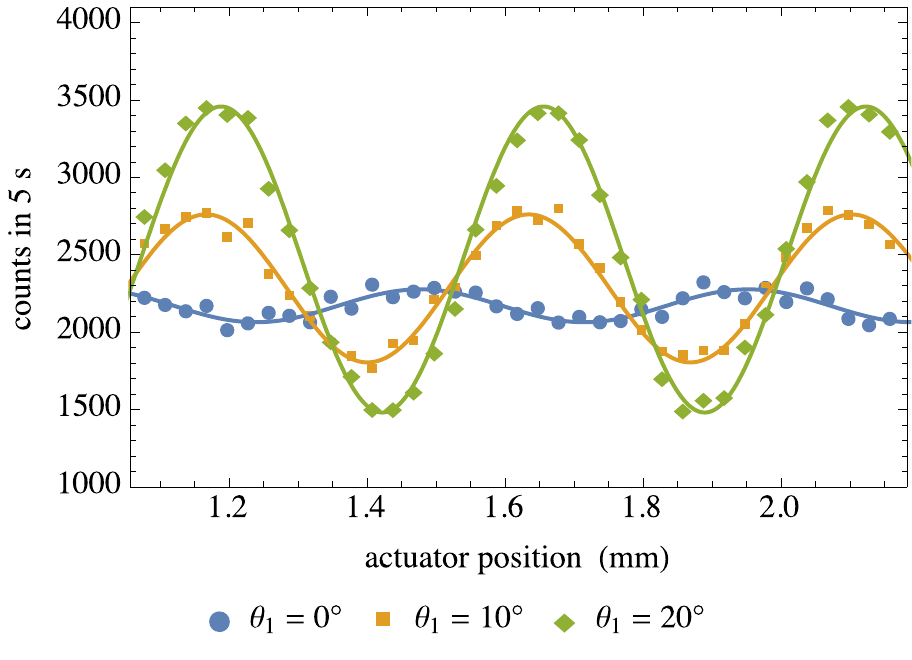}

\caption{\label{fig:rots}(Color online) Photon counts (per \unit[5]{s}) when a weak measurement of circular polarization is performed in (a) arm~2 and (b) arm~1 of the interferometer, shown as a function of actuator position (relative phase). The measurement is implemented through a small polarization rotation by an angle $\theta_k$ in arm $k$. Solid lines are sinusoidal fits. In agreement with theoretical predictions, the visibility of the interference pattern is substantially affected only by the measurement in arm~1. }
\end{figure}

Next, we perform weak measurements of circular polarization without testing for presence ($T_1=T_2=1$). We accomplish this by removing the Brewster's-angle glass slide and introducing a small polarization rotation $\theta_k$ in arm~$k$ ($k=1,2$). The resulting photon counts for three different polarization rotation angles ($\theta_k=0^\circ, 10^\circ, 20^\circ$) are shown in Fig.~\ref{fig:rots}. 

As expected, a polarization rotation in arm~2 [Fig.~\ref{fig:rots}(a)] has no noticeable effect on the visibility of the residual interference fringes (the visibility is around 2\% at all three angles $\theta_2=0^\circ, 10^\circ, 20^\circ$). At the same time, the overall intensity decreases with $\theta_2$, because increasing $\theta_2$ reduces the amplitude of the horizontal polarization component in arm~2, which is the postselected component. Within the experimental error dominated by the $2^\circ$ uncertainty in polarization rotation, the observed intensity decrease agrees with the $\cos^2\theta_2$ dependence predicted by Eq.~\eqref{eq:wkpreditt}. While this decrease represents a measurable influence of the polarization measurement on the detection probability, it is of second order in $\theta$ and is therefore neglected within the context of the quantum Cheshire cat, which relies on the assumption of weak measurement. Indeed, the intensity at $\theta_2=20^\circ$ differs by less than 4\% from the intensity in the absence of a polarization measurement ($\theta_2=0^\circ$).

When the rotation is instead made in arm~1 [Fig.~\ref{fig:rots}(b)], we observe a substantial increase in the visibility of the interference pattern with rotation angle $\theta_1$, as also expected from Eq.~\eqref{eq:visvis}. This demonstrates the second part of the quantum Cheshire cat. We calculate the theoretically predicted visibilities from Eq.~\eqref{eq:visvis} with $T_1=T_2=0$ and $\theta_2=0$. Since in our experiment the maximum observed visibility (at $\theta_1\approx 90^\circ$) is $V_m=0.72$, we scale the calculated visibilities by this factor. Then the predicted visibilities are  $0.24$ at $\theta_1=10^\circ$ and $0.44$ at $\theta_1=20^\circ$. We use the same calculation to estimate the uncertainty $\delta V=0.05$ in visibility due to the $2^\circ$ uncertainty in polarization rotation. Then the measured fringe visibilities obtained from sinusoidal fits of the data are $V_1(10^\circ)=0.21(5)$ and $V_1(20^\circ)=0.40(5)$, with a residual visibility of $\approx$5\% at $\theta_1=0^\circ$ caused by experimental imperfections in the polarization settings. The measured visibilities agree with the predicted values within the error. Figure~\ref{fig:rots}(b) also shows that the intensity increases with $\theta_1$. This is expected, since rotating the polarization of arm~1 increases the amplitude of the horizontally polarized component in that arm, with subsequent postselection of horizontal polarization; see also Eq.~\eqref{eq:wkpreditt}, which predicts the intensity to be proportional to $1+\sin^2\theta_1$.

To determine the experimental weak values, we note that for the ideal pre- and postselected states given by Eqs.~\eqref{eq:hvsguhtddhgu} and \eqref{eq:post}, one has $\text{Im}\, \langle \op{\sigma}_\text{circ}\op{\Pi}_1\rangle_w = -\cos\phi$. Therefore, the imaginary part of the weak value $\langle \op{\sigma}_\text{circ}\op{\Pi}_1  \rangle_w$ exhibits a sinusoidal dependence on $\phi$ [see Eq.~\eqref{eq:wk}], representing interference fringes for $\theta_1 \not= 0$ [see Eq.~\eqref{eq:p3}]. This sinusoidal dependence also holds for the generalized  pre- and postselected states given by Eqs.~\eqref{eq:ttt1} and \eqref{eq:ttt2}. We thus write $\text{Im}\, \langle \op{\sigma}_\text{circ}\op{\Pi}_k\rangle_w = -\abs{ \langle \op{\sigma}_\text{circ}\op{\Pi}_k\rangle_w} \cos\phi$ and use Eq.~\eqref{eq:hyhyhy} to relate the magnitude $\abs{ \langle \op{\sigma}_\text{circ}\op{\Pi}_k\rangle_w}$ of the weak value to the measured fringe visibility $V_k(\theta)$ when the polarization is rotated by $\theta$ in arm~$k$ \cite{Note1}. Scaling those visibilities by $V_m$ and using the measured weak values $\langle \op{\Pi}_k\rangle_w$, the experimental weak values (averaged over both settings $\theta_1=10^\circ$ and $\theta_1=20^\circ$) are $\abs{ \langle \op{\sigma}_\text{circ}\op{\Pi}_1\rangle_w}=0.86(21)$ and $\abs{ \langle \op{\sigma}_\text{circ}\op{\Pi}_2\rangle_w}=0.06(20)$, which agree within the error with the predictions $\abs{ \langle \op{\sigma}_\text{circ}\op{\Pi}_1\rangle_w}=1$ and $\abs{ \langle \op{\sigma}_\text{circ}\op{\Pi}_2\rangle_w}=0$ given by  Eq.~\eqref{eq:wk}. Note that to lowest order in $\theta$, Eq.~\eqref{eq:hyhyhy} gives $\abs{ \langle \op{\sigma}_\text{circ}\op{\Pi}_k\rangle_w} = (2\theta_k)^{-1} V_k(\theta)$. Since $2\theta_k$ is the predicted visibility for small $\theta$ [see Eq.~\eqref{eq:kjdvghdv}], this relation shows that in this limit $\abs{ \langle \op{\sigma}_\text{circ}\op{\Pi}_k\rangle_w}$ represents the ratio of measured to predicted visibility, analogously to Eq.~\eqref{eq:aahvsguhthgu222}.

\subsection{Simultaneous weak measurements of presence and polarization}

\begin{figure}

\begin{flushleft}
\hspace{0cm}{\small (a)}
\end{flushleft}

\vspace{-.7cm}

\includegraphics[scale=.9]{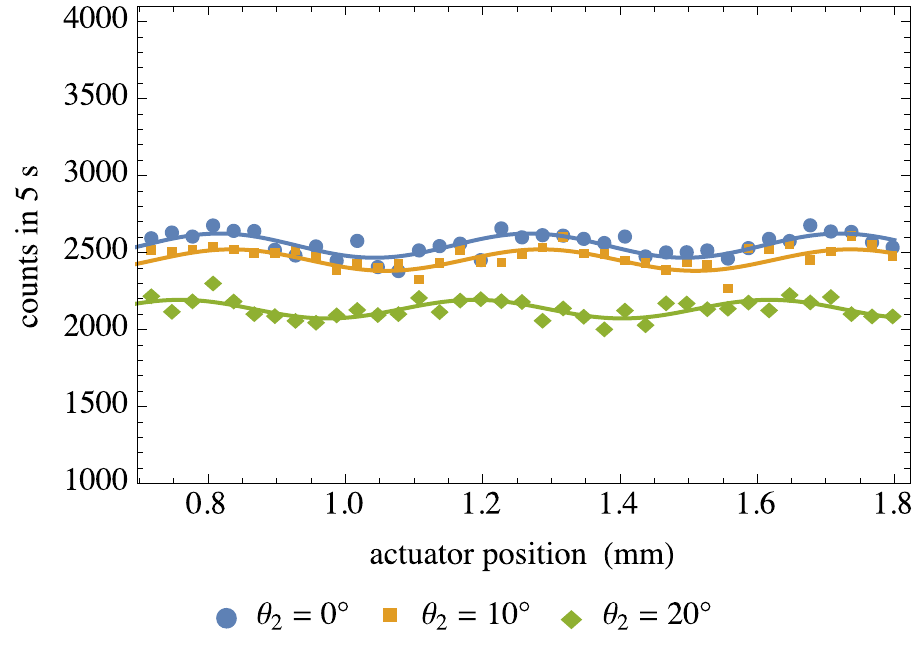}

\vspace{-.3cm}

\begin{flushleft}
\hspace{0cm}{\small (b)}
\end{flushleft}

\vspace{-.7cm}

\includegraphics[scale=.9]{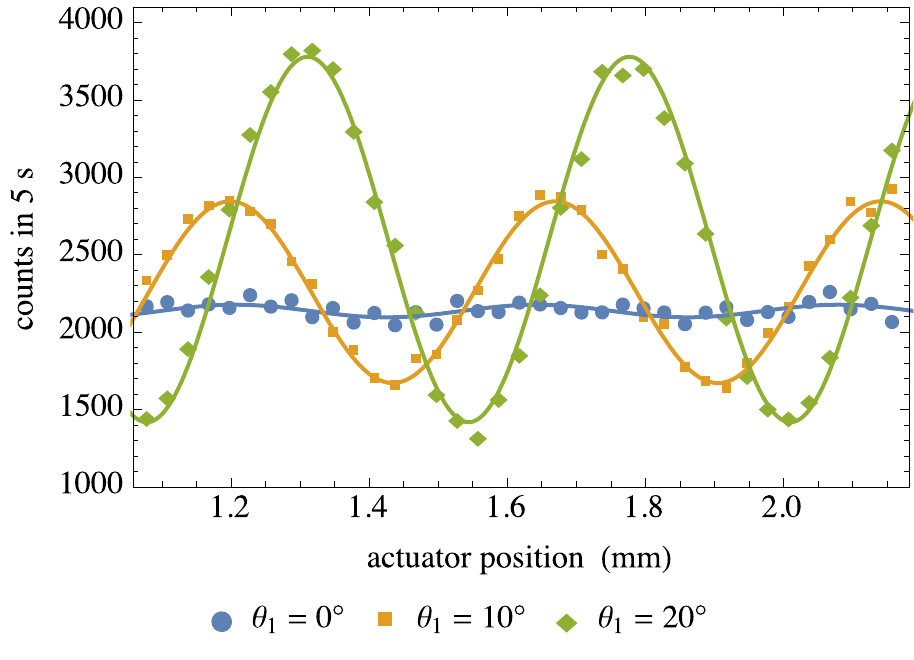}

\caption{\label{fig:simul}(Color online) Photon counts (per \unit[5]{s}) for simultaneous weak measurements of presence and circular polarization, shown as a function of actuator position (relative phase). Solid lines are sinusoidal fits. (a) Presence measurement in arm~1 and polarization measurement in arm~2. (b) Presence measurement in arm~2 and polarization measurement in arm~1.}
\end{figure}

Finally, we perform simultaneous weak measurements of presence in one arm and circular polarization in the other arm. Figure~\ref{fig:simul}(a) shows the results when a measurement of presence is performed in arm~1 and a polarization measurement is performed in arm~2. Just as for the polarization measurement in the absence of filtering (see Sec.~\ref{sec:weak-meas-polar}), the polarization measurement does not affect the residual fringe visibility, which remains around 3\%. Within the error, the average count rates agree with the count rates without filter at each setting of $\theta_2$ shown in Fig.~\ref{fig:rots}(a) (less than 4\% difference), demonstrating that the intensity is not substantially affected by the presence of the filter in arm~1. As discussed in  Sec.~\ref{sec:weak-meas-polar}, the polarization rotation also leads to an intensity decrease due to the reduction in the amplitude of the horizontal polarization component in arm~2, but this effect is of second order in $\theta_2$ and therefore not relevant within the weak-measurement approximation underlying the quantum Cheshire cat.

Figure~\ref{fig:simul}(b) shows the results when a polarization measurement is performed in arm~1 and a measurement of presence in arm~2. As already seen in Sec.~\ref{sec:weak-meas-polar}, the  polarization measurement has significant influence on the fringe visibilities. The measured visibilities are $V_1(10^\circ)=0.26(5)$ and $V_1(20^\circ)=0.45(5)$, with a residual visibility of 2\% at $\theta_1=0^\circ$. These values agree within the error with the predicted visibilities of $0.26$ at $\theta_1=10^\circ$ and $0.47$ at $\theta_1=20^\circ$. These visibilities are higher than those measured in the absence of the filter [see Fig.~\ref{fig:rots}(b)]. This is so because the reduction of the amplitude in arm~2 means that the amplitudes of the interfering horizontal components from both arms become more similar [see also Eq.~\eqref{eq:visvisrotabs}]. The measured intensity at $\theta_1=0$ becomes reduced by $15.4(8)$\% when the filter is placed in arm~2 compared to the intensity at $\theta_2=0$ when arm~1 is filtered (in agreement with the predicted value of $R=14.8$\%), indicating sensitivity of arm~2 to a measurement of presence. Note that the intensity increases with $\theta_1$ due to the increase of the amplitude of the horizontal polarization component in arm~1. 

Thus, we have demonstrated a single-photon quantum Cheshire cat also for simultaneous weak measurements of presence and circular polarization. On average, these measurements have an effect only if the measurement of presence is carried out in arm~2 and the simultaneous measurement of polarization is carried out in arm~1.

\section{\label{sec:conclusion}Conclusion}

Our experiment realizes an all-optical, single-photon quantum Cheshire cat. As predicted by both the weak-value and wave-function formalisms, whether the observed photon counts are sensitive to weak measurements of photon presence and circular polarization depends on the particular arm of the interferometer in which each measurement is carried out. We find that, on average and for the pre- and postselected states used in the experiment, photons passing through one of the arms are not appreciably affected by weak measurements of presence, while photons in the other arm are not appreciably affected by weak measurements of polarization. We explicitly confirm this result also when the two measurements are carried out simultaneously.

This behavior is succinctly represented by the weak values for these two kinds of weak measurements: The theoretical prediction for the weak value in one of the arms is zero while it has unit magnitude in the other arm. We measured these weak values from the change in the observed photon counts when the weak measurements are introduced and found good agreement with the theoretical values. Dominant sources of error in our experiment are imperfections in the preparation, manipulation, and postselection of the polarization states. 

The extent to which one considers a quantum Cheshire cat to establish a paradoxical situation depends largely on how one interprets the physical meaning of weak values, and several such interpretations have been suggested \cite{Aharonov:1988:mz,Duck:1989:uu,Johansen:2004:ll,Aharonov:2005:oo,Svensson:2014:xx,Dressel:2014:uu}. Our experiment emphasizes that weak values are ensemble averages \cite{Note2}, because the weak values are measured from changes in the detected intensities and fringe visibilities, requiring accumulation of many photon events. As has been shown here and elsewhere  \cite{Atherton:2015:oo,Correa:2015:ii}, the observed measurement statistics for a quantum Cheshire cat can also be understood as a consequence of ordinary quantum interference, without recourse to weak values. 

Quantum Cheshire cats highlight peculiar features of weak values, just as the first work on weak values did \cite{Aharonov:1988:mz}. Our experiment provides a powerful and physically transparent implementation that illuminates these features and aids in the analysis and interpretation of the quantum Cheshire cat.

\begin{acknowledgments}
The authors are indebted to M.~Beck for helpful conversations and crucial equipment advice. This research was supported by the Foundational Questions Institute and the Student Summer Science Scholar program of the University of Portland.
\end{acknowledgments}

% \bibliography{../Bib/References} 
% \bibliographystyle{apsrev4-1}


\end{document}